\documentclass[aps,prl,twocolumn,groupedaddress,showpacs,10pt]{revtex4-1}
\usepackage{graphicx}
\usepackage{amsfonts}
\usepackage{color}

\begin{document}

%\preprint{}

% Updated: 26 Mar 2018

\title{Deflection of a Molecular Beam Using the Bichromatic Stimulated Force}

\author{S. E. Galica}
\author{L. Aldridge}
\altaffiliation[Present address: ]{Department of Physics, Yale University, New Haven, CT 06520, USA.}
\author{D. J. McCarron}
\author{E. E. Eyler} 
\thanks{Deceased}
\author{P. L. Gould}
\affiliation{Physics Department, University of Connecticut, Storrs, CT 06269, USA}

\date{\today}

\begin{abstract}

We demonstrate that a bichromatic standing-wave laser field can exert a significantly larger force on a molecule than ordinary radiation pressure. Our experiment measures the deflection of a pulsed supersonic beam of CaF molecules by a two-frequency laser field detuned symmetrically about resonance with the nearly closed $X (v=0) \rightarrow B (v'=0)$ transition. The inferred force as a function of relative phase between the two counterpropagating beams is in reasonable agreement with numerical simulations of the bichromatic force in this multilevel system. The large magnitude of the force, coupled with the reduced rate of spontaneous emission, indicates its potential utility in the production and manipulation of ultracold molecules.
\end{abstract}


\maketitle

% body of paper here - No section commands for PRL

Radiative forces on atoms have been the major tool enabling laser cooling and trapping \cite{Metcalf1999} and the myriad of applications which have resulted, including precision spectroscopy, quantum degenerate gases, ultracold collisions, and quantum simulations. There are two general types of radiative force: spontaneous force, also known as radiation pressure, and stimulated force, also known as the dipole force. Radiation pressure is the result of absorption/spontaneous emission cycles, while the stimulated force arises from absorption followed by stimulated emission. The latter requires an intensity gradient and can be thought of as a coherent redistribution of photons between various propagation directions. Laser manipulation of atoms is a well-developed field, but recently, these techniques have been increasingly applied to molecules \cite{Bohn2017}. This extension is nontrivial due to the complicated internal structure of molecules caused by their vibrational and rotational degrees of freedom \cite{DiRosa2004}. Radiation pressure has been used to slow, cool, and trap molecules that fortuitously have near-cycling transitions \cite{Barry2012,Hemmerling2016,Truppe2017,Hummon2013,Barry2014,Norrgard2016,Anderegg2017,Truppe2017a,Kozyryev2017,Lim2018}. Stimulated forces have also been used to manipulate molecules \cite{Hill1975,Voitsekhovich1994,Stapelfeldt1997,Bishop2010}, but on a more limited scale. Compared to radiation pressure, stimulated forces have two significant advantages for molecules: (1) radiation pressure is limited by the spontaneous emission rate, while stimulated forces can greatly exceed this saturated value; and (2) radiation pressure relies on spontaneous emission which can optically pump the molecules into ``dark" states which no longer interact with the laser field.

A specific type of stimulated force, the bichromatic force (BCF) \cite{Soding1997,Metcalf2017}, is particularly promising for manipulating molecules \cite{Chieda2011}. The BCF involves two frequencies which are tuned symmetrically above and below a resonant frequency $\omega$ by $\pm\delta$. The two frequencies are both present in oppositely-directed beams, which gives rise to counterpropagating trains of beat notes with a fixed relative phase, $\chi$. In a simplified picture of BCF, if each beat is considered an effective $\pi$-pulse which inverts the population, a molecule can be excited by a beat from one direction and rapidly returned to the ground state by a beat from the other direction. Each absorption/stimulated emission cycle imparts an impulse of $2\hbar k$ and repeats at a rate determined by the beat frequency, which is set by the detuning, $\delta$. For a given detuning, the Rabi frequency must be maintained at $\Omega_r=\sqrt{3/2}\delta$ to maintain the effective $\pi$-pulse condition and properly cycle the population. The rate of momentum transfer, i.e. force, can thus increase arbitrarily with detuning, provided there is sufficient laser intensity to maintain the correct Rabi frequency. The direction of force is controlled by $\chi$, since this determines the sequence of beat-pulse arrivals. There are several beneficial aspects of BCF: it can be much larger than the radiation pressure force; it has a large range of velocities over which it is relatively constant; and it does not rely on spontaneous emission thus does not inherently suffer from the problem of optical pumping into dark states. All of these are important for slowing a beam of molecules. The large force means that the beam can be slowed over a short distance, resulting in less spreading and higher brightness. The large capture range means that higher velocities and broader velocity distributions can be slowed. The relative unimportance of spontaneous emission reduces the need for multiple lasers to retrieve population lost to optical pumping. This makes BCF more applicable than the spontaneous force in manipulating diatomic and polyatomic molecules which lack near-cycling transitions due to poor vibrational overlap.

\begin{figure}
	\includegraphics[width=\linewidth]{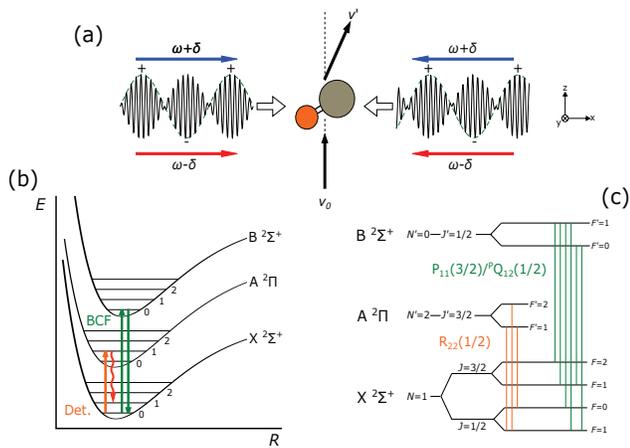}
	\caption{(a) Scheme for measurement of the bichromatic force (BCF). The CaF molecular beam experiences a transverse deflection by interacting with a bichromatic standing wave. (b) Relevant molecular potentials and transitions \cite{Chieda2011,Aldridge2016} in CaF. For BCF: $X (v=0) \rightarrow B (v'=0)$, and for fluorescence detection: $X (v=0) \rightarrow A (v'=1)$. Energy spacings are not to scale. (c) Expanded view of BCF and detection transitions showing fine ($J$) and hyperfine ($F$) structure for selected rotational ($N$) levels. The hyperfine states in $X$ span 146 MHz, while the hyperfine splittings in $A$ and $B$ are negligible \cite{Tarbutt2015}. The BCF uses the combined $P_{11}(3/2)/^PQ_{12}(1/2)$ transition, while detection employs $R_{22}(1/2)$.}
	\label{fig:bcf_cartoon}
\end{figure}

BCF has been observed and studied in various atomic systems \cite{Soding1997,Williams1999,Cashen2001,Partlow2004,Liebisch2012,Chieda2012,Feng2017}. Here we demonstrate the effectiveness of BCF in a molecular system via the transverse deflection of a beam of CaF, a polar diatomic molecule, as shown in Fig. \ref{fig:bcf_cartoon}(a). We find the force exhibits the expected dependence on the relative phase, $\chi$, and is larger, at our achieved laser intensity, than the maximum radiation pressure force by a factor of $\sim$4. Significantly, the deflection does not require measures to counteract rovibrational optical pumping. Overall, the measurements are in reasonable agreement with numerical simulations of the multilevel molecular system. We note that similar results on BCF have been recently reported with triatomic SrOH molecules \cite{Kozyryev2018}, a system which required an additional laser for rovibrational repumping.

\begin{figure}
	\includegraphics[width=\linewidth]{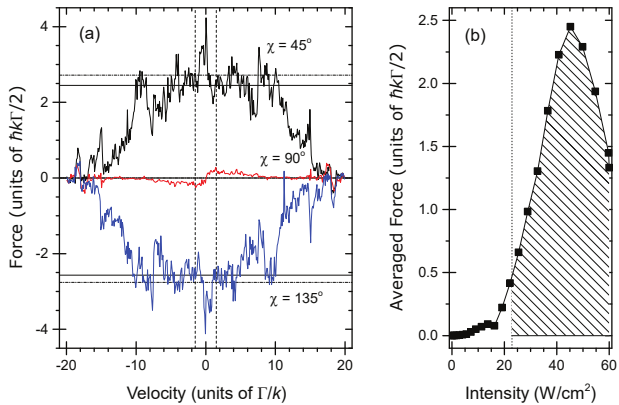}
	\caption{(a) BCF vs. $v$ for various phases, $\chi$, for $\delta=30\Gamma=2\pi(190$ MHz) and $\Omega_r=\sqrt{3/2}\delta=2\pi(232.7$ MHz). Vertical dashed lines indicate the range of velocities present in our molecular beam. Horizontal lines indicate the average force over this range using a straight average (dashed lines), and one where the sharp upward spikes have been omitted (solid lines), as described in the text. (b) Average force for $\chi=45^o$, with spikes omitted, vs. intensity. The shaded region denotes the range of peak intensities corresponding to the vertical $(y)$ extent of our molecular beam. The Rabi frequency is computed as $\Omega_r=2\pi\sqrt{I/3}(60$ MHz) where $I$ is in units of W/cm$^2$ (see \cite{Aldridge2016T}).}
	\label{fig:BCF_profiles}
\end{figure}

The basic theory of BCF in a two-level system has been dealt with from various perspectives, including the $\pi$-pulse picture \cite{Voitsekhovich1989,Voitsekhovich1991,Soding1997,Cashen2003}, numerical simulations of the optical Bloch equations \cite{Williams1999}, and the doubly-dressed-atom picture \cite{Grimm1994,Yatsenko2004}. Extensions to multilevel systems, which are necessary for applying BCF to molecules, have also been carried out \cite{Aldridge2016,Yang2017,Yin2018}. We have recently reported \cite{Aldridge2016} on such numerical calculations of the time-dependent density matrix for CaF, which we briefly summarize here. The BCF is applied using the $X (v=0) \rightarrow B (v'=0)$ branch at 531 nm on the $P_{11}(3/2)/^PQ_{12}(1/2)$ rotational transition, as shown in Figs. \ref{fig:bcf_cartoon}(b) and \ref{fig:bcf_cartoon}(c). This transition is nearly closed with respect to vibration---Franck-Condon factor (FCF) = 0.9987---and selection rules prevent loss to other rotational levels. The excited-state decay rate is $\Gamma = 2\pi(6.34$ MHz). The resonant frequency, $\omega$, is taken to be between the degeneracy-weighted locations of the excited-state and ground-state hyperfine manifolds. The former splitting is negligible \cite{Tarbutt2015}, while the latter extends over 146 MHz. As verified in the calculations \cite{Aldridge2016}, this hyperfine structure does not require sidebands for repumping due to the relatively large Rabi frequencies and detunings used. The linearly-polarized electric field has four components of equal amplitude, $E_0$: two counterpropagating beams (along $\pm x$), each containing two frequencies, $\omega\pm\delta$. The phase difference between the electric fields of the counterpropagating beat notes is $\chi$. We incorporate a total of 16 states, which includes the rotation, fine structure, and hyperfine structure with the associated magnetic sublevels. Rabi frequencies are defined for each possible transition between ground and excited states, and spontaneous decay is included. To prevent population from being trapped in dark sublevels, a skewed magnetic field of magnitude 29.2 G oriented at $71^o$ with respect to the laser polarization, as employed in the experiment, is included as well. The transverse $(x)$ velocity enters via the substitution $x=vt$ in the expression for the electric field, and the velocity-dependent force is obtained from the density matrix elements by applying Ehrenfest's theorem. Examples of BCF vs. velocity for various phases, $\chi$, are shown in Fig. \ref{fig:BCF_profiles}(a). Of particular note is the enhancement of BCF relative to the saturated radiation pressure force, $\hbar k\Gamma/2$, as well as the broad range of velocities over which BCF is relatively constant. The dependence of the velocity-averaged force on $\chi$ is striking: it is maximized for $\chi=45^o$, effectively vanishes for $\chi=90^o$, and is inverted for $\chi=135^o$. Care must be taken in calculating the velocity-averaged force. The force vs. velocity profile exhibits many narrow spikes which should not be included because such large forces act only for a short time before the velocity change pushes the molecule off the spike. The averaging which accounts for this effect is done by calculating the mean and standard deviation, $\sigma$, for the distribution and identifying those points whose magnitude \textit{exceeds} the mean by $>\sigma$. These points are reset to the mean, and the process is repeated until convergence is achieved. Points whose magnitude is less than the mean are not adjusted since these less-than-average forces result in small velocity changes which can act for extended periods. We note that excluding the spikes in this way typically reduces the average force by $<10\%$, as shown in Fig. \ref{fig:BCF_profiles}(a).

In the experiment, we subject a supersonic beam of CaF to BCF and measure the force via transverse deflection of the beam. The pulsed molecular beam is generated by ablating a Ca rod situated in front of a 1 mm diameter nozzle from which a mixture of Ar (98\%) and SF$_6$ (2\%) at a backing pressure of 30 psig emerges. The ablation employs a 10 ns, 10 mJ pulse from a frequency-doubled Nd:YAG laser focused to $\sim$1 mm diameter. The rod is slowly rotated and translated to reduce shot-to-shot signal variations. Various species result from the ablation and subsequent reactions, one of which is the CaF of interest. A pulsed valve (Lasertechnics LPV piezo-electric valve) reduces the gas load into the chamber and yields a pulse of molecules with average longitudinal velocity $v_z=415$ m/s and a FWHM of 75 m/s, as measured by time-of-flight. A 1 mm diameter aperture located 131 mm downstream from the nozzle collimates the beam to 13.7 mrad (FWHM).

The BCF is applied in an interaction region 251 mm from the source. The light for the BCF derives from a distributed-Bragg-reflector (DBR) diode laser at 1062 nm which is amplified in a fiber amplifier then frequency doubled in a single pass through a periodically-poled lithium niobate crystal, providing a maximum of 1.3 W at 531 nm. The two frequencies for the BCF, $\omega\pm\delta$ with $\delta=30\Gamma=2\pi(190$ MHz), are generated by two acousto-optical modulators (AOMs) and combined on a beamsplitter to provide matched intensities. At this detuning, the peak intensity is 60 W/cm$^2$ in each beam frequency component, corresponding to a Rabi frequency of $\Omega_r=2\pi(268$ MHz) \cite{Aldridge2016T}. Light at the central frequency, $\omega$, is used in an I$_2$ saturated absorption spectrometer to lock the frequency of the DBR laser. The two-frequency beam is focused to an elliptical FWHM of 0.82 mm horizontally by 0.56 mm vertically at its perpendicular intersection with the molecular beam. The counterpropagating two-frequency beam is provided by retroreflection from a plane mirror outside the vacuum chamber. The distance of this mirror from the molecular beam determines the phase difference, $\chi$. For our detuning $\delta=2\pi(190$ MHz), $\chi=45^o$ is obtained for a distance of 98.6 mm. For larger phase differences, the intensity imbalance due to the expansion of the retroreflected beam must be accounted for. The interaction time is sufficiently short that optical pumping out of the nearly-closed transition is negligible.

The deflection caused by the BCF is measured by fluorescence. A distance 223 mm downstream from the interaction region, light from a cw dye laser perpendicularly (along $x$) intersects the molecular beam immediately in front of a 0.76 mm slit which is scanned along $x$ in front of a fixed PMT (Hamamatsu H10721-20) detector. The resulting fluorescence is collected and collimated by a 20 mm focal-length, 25 mm diameter asphere and imaged by a matching lens onto the detector's 8 mm diameter active area. The vertical ($y$) aperture of the detector is restricted to reduce both scattered light from the walls of the vacuum chamber and signal from molecules not passing through the high-intensity region of the BCF beams. The detection laser is locked to the $X (v=0) \rightarrow A (v'=1)$, $R_{22}(1/2)$ transition at 583 nm using I$_2$ saturated absorption. Although this non-diagonal transition has a poor FCF$\sim0.013$, it is saturated with the utilized power of 300 mW focused to an elliptical FWHM of 0.47 mm horizontally by 1.41 mm vertically. This transition has the advantage that the strongest fluorescence, $A (v'=1) \rightarrow X (v''=1)$, is separated in wavelength by $\sim$23 nm from the excitation transition, allowing background scattered light to be reduced via filtering. The tradeoff is that each molecule emits only one photon before populating the dark $X(v''=1)$ state.

Data are obtained by integrating the fluorescent signal during a 675 $\mu$s gate centered on the arrival time of the peak of the molecular beam, $\sim1$ ms after the pulsed valve opens. A second gate, centered 2.25 ms after the molecular peak, is used to measure the background scattered light which is then subtracted from the signal occurring in the first gate. For each deflection condition, we conduct typically 9 trials. In each, the slit position $x$ is varied randomly, and at each position, 5000 shots are collected, each yielding $\sim$2.5 photons. To reduce systematic errors, this is repeated for a different random sequence of slit positions. In total, 90,000 shots are typically collected at each value of $x$. Error bars are statistical, dominated by shot-to-shot fluctuations in the molecular source.

\begin{figure}
	\includegraphics[width=\linewidth]{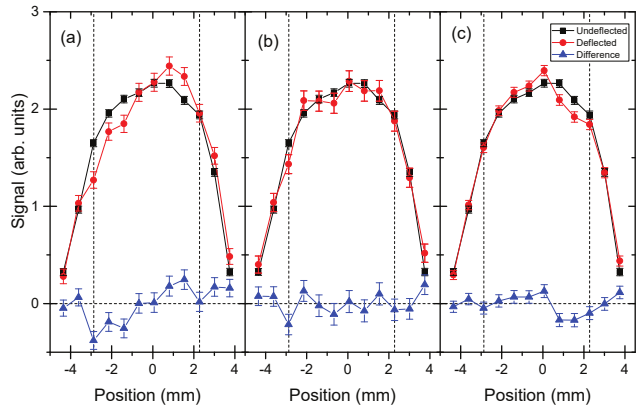}
	\caption{Deflection profiles and differences for various phases: (a) $\chi=45^o$, (b) $\chi=90^o$, (c) $\chi=135^o$; all at detuning $\delta=30\Gamma$ and peak intensity 60 W/cm$^2$. The dashed vertical lines denote the range used for calculation of deflections.}
	\label{fig:data}
\end{figure}

\begin{figure}
	\includegraphics[width=\linewidth]{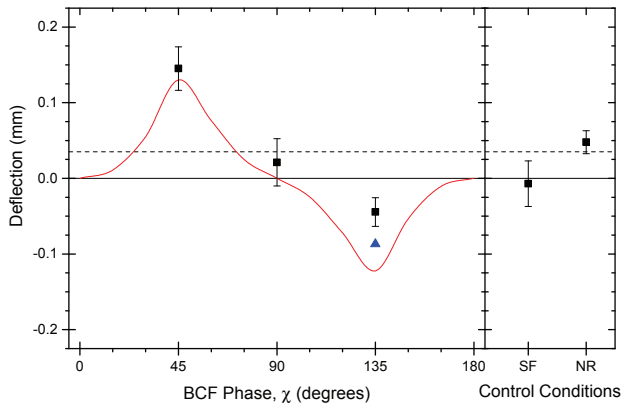}
	\caption{Deflection vs. phase, $\chi$. Experimental points (squares) are shown for $\chi=45^o$, $90^o$, and $135^o$. Also shown is a simulation (solid curve), with no adjustable parameters, which includes the averaging described in the text, but does not include intensity imbalance. For the point at $\chi=135^o$, where the intensity imbalance is most significant, we include a simulated point (triangle) where the actual $y$ and $z$ dependencies of the imbalance are explicitly accounted for, as described in the text. The horizontal dashed line is the simulated deflection for an intensity imbalance of 100\%, corresponding to the radiation pressure from the bichromatic field. Also shown, at right, are measured deflections when there is no retroreflected beam (NR) and a single-frequency standing wave (SF).}
	\label{fig:defPhase}
\end{figure}

Scans of the undeflected and deflected beams for various phases, $\chi$, are shown in Fig. \ref{fig:data}. Also shown are the differences between the two curves, which clearly indicate the directions of the force: $+x$ for $\chi=45^o$, 0 for $\chi=90^o$, and $-x$ for $\chi=135^o$. To compare with theory, we calculate the deflection as the difference between the average values of $x$ for the deflected and undeflected curves. Because of uncertainties in the fluorescence collection efficiency in the wings of the molecular beam profiles, the deflection calculations only include the central regions, indicated by the dashed vertical lines. Deflection vs. $\chi$ is plotted in Fig. \ref{fig:defPhase}. Error bars are statistical, resulting from the propagation of uncertainties in the data points of Fig. \ref{fig:data} through the deflection calculation.

To compare the measured deflections with the results of numerical simulations, several factors must be accounted for. The BCF has a significant dependence on intensity, as seen in Fig. \ref{fig:BCF_profiles}(b). The aperture on the detector has a vertical ($y$) extent of 1.5 mm, which extrapolated back to the interaction region, corresponds to 0.79 mm. Since the incident elliptical BCF beam has a FWHM of 0.56 mm along $y$, not all molecules detected at a given value of $x$ will have experienced the same intensity. Furthermore, as a molecule propagates along $z$ through the BCF beams, it experiences a time-varying BCF intensity. For a given value of $y$, we calculate the $x$-momentum transfer by integrating the intensity-dependent BCF force, $F$, over time $t=v_z/z$: $\Delta p_x(y) = \int F(y,z) dz/v_z$. Since the transverse spread of velocities in the undeflected beam is significantly less than the range of velocities affected by BCF (see Fig. \ref{fig:BCF_profiles}(a)), we use an average value of the force over the relevant velocity range, as discussed earlier and shown in Fig. \ref{fig:BCF_profiles}(b). We repeat this for various values of $y$ and average vertically over the detector aperture. This $h$ = 1.5 mm aperture is sufficiently narrow that the undeflected beam profile is taken to be independent of $y$: $I_0(x,y) = Af(x)$, where $f(x)$ is the undeflected curve in Fig. \ref{fig:data}. The signal corresponding to the deflected beam profile $I(x,y)$ is given by: 
\begin{equation}
S(x)=\int_{-h/2}^{+h/2}I(x,y)dy=A\int_{-h/2}^{+h/2}f(x-\Delta p_x\tau/m)dy,
\end{equation}
where m is the molecular mass and $\tau$ is the time-of-flight from the interaction region to the detection region. From this calculated signal, $S(x)$, we can extract the average deflection $x_d$. As with the experimental data, we only use the central region of the profiles for these calculations. These average values of $x_d$ are shown as a function of $\chi$, together with the measured values, in Fig. \ref{fig:defPhase}. We show two versions of the calculations. In the first (solid curve), we assume the two counterpropagating beams have matched intensities. In the second, shown only for $\chi=135^o$, we explicitly account for the $y$ and $z$ dependencies of the imbalance, obtained from measured beam profiles, when calculating the force. This is nontrivial because of the ellipticity of the BCF beams and imbalance reversals in certain regions. This more realistic accounting for imbalance shows better agreement with the $\chi=135^o$ data point, but still overestimates the deflection by 2.2 standard deviations. This could be due to imperfect overlap of the BCF beams and/or deflection due to radiation pressure. We also calculated the deflection in the extreme case of 100\% intensity imbalance, i.e. no retroreflected intensity, shown as the dashed horizontal line. This corresponds to radiation pressure and demonstrates that the BCF can provide a significantly larger force.

We have conducted control experiments, also shown in Fig. \ref{fig:defPhase}, to confirm that the observed deflections are indeed due to BCF. First, we blocked the retroreflected beam, which results in radiation-pressure deflection. The measured deflection is consistent with that calculated from radiation pressure and is significantly less than deflection from the maximum BCF. We then restored the retroreflected beam (at $\chi=45^o$), but blocked the $\omega+\delta$ frequency, yielding a simple standing wave of frequency $\omega-\delta$. The deflection in this case is consistent with zero, as expected.

In conclusion, we have observed the action of the bichromatic force (BCF) on molecules by measuring the deflection of a beam of CaF. The inferred force is significantly larger than the ordinary radiation pressure force, demonstrating that this stimulated force, with its broad velocity capture range and non-reliance on spontaneous emission, will be a welcome addition to the arsenal of molecule manipulation techniques. It will be particularly useful in rapidly decelerating molecular beams for efficient trap loading, especially if chirping is incorporated to compensate for the changing Doppler shift \cite{Chieda2012}. For such applications, the use of a cryogenic buffer-gas-cooled molecular beam \cite{Lu2011,Kozyryev2018} would be beneficial. If the BCF parameters used here were applied to such a CaF beam ($v_{initial}$ = 60 m/s), stopping could be achieved in $\sim$2 cm, 4x shorter than if the maximum radiation pressure force were used. This would yield a 16x brighter beam for trap loading. Even without a repumper laser, 27\% of the population would remain in $v = 0$ \cite{Aldridge2016T}, compared to only 0.2\% when using radiation pressure. Further improvements in BCF are expected by utilizing higher detunings/intensities and employing polychromatic \cite{Galica2013} instead of bichromatic fields.

This work was supported by the National Science Foundation and the US-Israel Binational Science Foundation.

% \begin{figure}
% \end{figure}


\end{document}